\documentclass[twocolumn,aps,nofootinbib]{revtex4-1}
\usepackage{graphicx}
\usepackage{epstopdf}
\usepackage{latexsym}
\usepackage{amsmath}
\usepackage{amssymb}
\usepackage{xcolor}
\usepackage{mathrsfs}
\usepackage{comment}
\usepackage[center]{subfigure}

\begin{document}
\title{Gravitational wave cosmology in Einstein-scalar-Gauss-Bonnet gravity}
\date{\today}
\author{Jared Fier$^{1}$}
\email{Jared$\_$Fier@baylor.edu}
\author{Henry Han$^{2}$}
\email{Henry$\_$Han@baylor.edu}
\author{Bowen Li$^{1}$}
\email{Bowen$\_$Li@baylor.edu}
\author{Kai Lin$^{3}$}
\email{lk314159@hotmail.com}
\author{Shinji Mukohyama$^{4, 5}$}
\email{shinji.mukohyama@yukawa.kyoto-u.ac.jp}
\author{Anzhong Wang$^{1}$}
\email{Anzhong$\_$Wang@baylor.edu; corresponding author}
\affiliation{
$^{1}$ GCAP-CASPER, Department of Physics, Baylor University, One Bear Place $\#$97316, Waco, Texas, 76798-7316, USA\\
$^{2}$ The Laboratory of Data Science and Artificial Intelligence Innovation, Department of Computer Science, School of Engineering and Computer Science, Baylor University, Waco, TX  76798 USA\\
$^{3}$ Universidade Federal de Campina Grande, Campina Grande, PB 58429-900, Brasil\\
$^{4}$ Center for Gravitational Physics and Quantum
Information, Yukawa Institute for Theoretical Physics, Kyoto University, 606-8502, Kyoto, Japan \\
 $^{5}$ Kavli Institute for the Physics and Mathematics of the Universe (WPI), The University of Tokyo Institutes for Advanced Study,
 The University of Tokyo, Kashiwa, Chiba 277-8583, Japan
}

\begin{abstract}
In the framework of Einstein-scalar-Gauss-Bonnet (EsGB) gravity, we systematically study gravitational waves (GWs), first produced by remote compact astrophysical sources and then propagating through the flat homogeneous and isotropic Universe at cosmic distances before arriving at detectors. Assuming that the speed $c_T$ of the spin-2 graviton is the same as that of photons, we find explicitly the stability conditions of the theory and then obtain the severest observational constraint found so far. In particular, all these conditions and constraints are satisfied, provided that $0 \leq \alpha\dot{f}(\phi_0) \lesssim 8.97 \times 10^{-24}$ (km), where $\alpha{f}(\phi)$ denotes the coupling strength between the scalar field $\phi$ and the Gauss-Bonnet term, an over-dot represents the derivative with respect to the cosmic time, and $\phi_0$ is the present value of $\phi$.    The trajectories for both spin-2 and spin-0 gravitons and the amplitudes of GWs along the trajectories are explicitly obtained. The amplitude of a spin-2 GW is practically indistinguishable from that of GR, while the spin-0 GWs remain almost constant during radiation- and matter-dominated epochs, and in the dark energy-dominated epoch it is proportional to the physical distance between the source and the observer. A careful analysis shows that the latter is due to the assumption $c_T = 1$. When $c_T \not= 1$ to the extent that is consistent with the stability conditions and observational constraints, the above behavior disappears. 
\begin{flushright} {\footnotesize YITP-25-29, IPMU25-0008}  \end{flushright}
\end{abstract}

\maketitle

\section{Introduction}
\renewcommand{\theequation}{1.\arabic{equation}} \setcounter{equation}{0}

The first detection of the gravitational wave (GW) from the coalescence of two massive black holes (BHs) by the advanced Laser Interferometer Gravitational-Wave Observatory
(aLIGO) marked the beginning of a new era,  {\it the GW astronomy} \cite{Ref1}. Following this observation, soon more than 90 GW events were detected by the LIGO/Virgo/KAGRA scientific
collaboration \cite{GWTC1,GWTC2,GWTC3}, and many more are expected to come after the Run O4 mission is completed \cite{aLIGO}. The outbreak of interest on GWs and BHs has further gained momenta after the detection of the shadows of the M$87^*$ and Sgr A$^*$ supermassive BHs \cite{EHT19,EHT22}.

One of the remarkable observational results is the discovery that the mass of  an individual BH in these binary systems can be  much larger than what was previously expected, both theoretically and observationally \cite{Ref4,Ref5,Ref6}, leading to the proposal and refinement of various formation scenarios, see, for example, \cite{Ref7,Ref8,SSTY18,LIGO-Virgo20}, and references therein.
A consequence of this discovery is that the early inspiral phase may also be detectable by space-based observatories, such as LISA \cite{PAS17}, TianQin \cite{TianQin}, Taiji \cite{Taiji}, and DECIGO \cite{DECIGO},
for several years prior to their coalescence \cite{AS16, Moore15}. Multiple observations with different detectors at different frequencies of signals from the same source can provide excellent
opportunities to study the evolution of the binary in detail. Since different detectors observe at disjoint frequency bands, together they cover different evolutionary stages of a binary system.
Each stage of the evolution carries information about different physical aspects of the source. As a result, multi-band GW detections will provide  an unprecedented opportunity to test
different theories of gravity in the strong field regime \cite{Carson:2019kkh} and open a completely new window to study cosmology - {\em GW cosmology} \cite{Bailes21}.

It is remarkable to note that the space-based  detectors mentioned above, together with the current and forthcoming  ground-based ones, such as   Voyager \cite{Voyager},   the Einstein Telescope (ET) \cite{ET} and Cosmic Explorer (CE) \cite{CE},
 are able to detect GWs emitted from  such systems as far as the redshift is about $z \simeq 100$ \cite{HE19}, 
which will result in a variety of profound scientific consequences. In particular, GWs propagating over such long cosmic distances will carry valuable information not only about their sources, but also about  the detail of the cosmological expansion and inhomogeneities of the universe, whereby a completely new window to explore the universe by using GWs is opened, as so far our understanding of the universe almost all comes from observations of electromagnetic waves only (possibly with the important exceptions of  cosmic rays and neutrinos) \cite{LL09}.

Recently, in the framework of GR we studied GWs first produced by remote compact astrophysical sources and then propagating through the inhomogeneous universe at cosmic distances, and found the conditions that the back-reactions of GWs to the background can be negligible, so the linear perturbations are valid \cite{Fier21}.  To simplify the field equations, we showed that the spatial, traceless, and Lorentz gauge conditions can be imposed simultaneously, even when the background is not vacuum.  Applying the general formulas together with the geometrical optics approximation to such GWs, we found that they still move along null geodesics and its polarization bi-vector is parallel-transported, even when both the cosmological scalar and tensor perturbations are present. In addition, we also calculated the gravitational integrated Sachs-Wolfe effects, whereby the dependence of the amplitude, phase and luminosity distances of the GWs on these two kinds of perturbations are read out explicitly. This generalized the results obtained previously by Laguna et al \cite{LLSY10}, in which only cosmological scalar perturbations were assumed to be present. 

In this paper we shall consider Einstein-scalar-Gauss-Bonnet (EsGB) gravity and, as a first step, systematically study the propagation of GWs in the flat homogeneous and isotropic universe through cosmic distances before arriving at detectors. In comparison with GR, a distinguishable feature  of  EsGB gravity is that it contains both spin-0 and spin-2 gravitons. Therefore, by studying their propagation through the universe, we seek  observational evidences for  EsGB gravity. 

Specifically, the rest of the paper is organized as follows:  In Section II, after briefly reviewing EsGB gravity, we study the stability conditions of the theory and its observational constraints. In the case $c_T = 1$, we find that the stability conditions hold in all epochs of the evolution of the universe,  provided that the condition (\ref{CD1}) is satisfied. Where $c_T$ denotes the speed of the spin-2 gravitons. In the same section, we also consider the observational constraints and obtained the condition  (\ref{eq2.4uu}), which represents the severest constraints obtained so far. 

In Section III we first decompose the perturbations of $h_{\mu\nu}$, which represent GWs propagating over the homogeneous and isotropic universe into scalar and tensor perturbations, and then study the propagation of the spin-0 and spin-2 modes  by using the geometrical optics approximation. To carry out the studies analytically, we impose the condition $c_T = 1$ over the whole evolution of the universe, including the radiation-, matter- and dark energy-dominated epochs, and are able to integrate the trajectories of both spin-2 and spin-0 gravitons explicitly, so are the amplitudes of GWs along the trajectories. We find that  the amplitude of the spin-2 GW is practically indistinguishable from that of GR, while the spin-0 GWs remain almost constant during radiation- and matter-dominated epochs, and in the dark energy-dominated epoch it is proportional to the physical distance between the source and observer. 

In Section IV, we study the propagation of the spin-0 graviton carefully without imposing the condition $c_T = 1$ in the dark energy-dominated epoch, but still assuming that the stability conditions and observational constraints are satisfied. Then, we show that the above behavior disappears, and instead the amplitude of the spin-0 GWs is given by Eq.(\ref{eq5.10}),
which is proportional to $\dot{f}(t) \left(= f_{,\phi} \dot{\phi}\right) \ll 1$.

The paper is ended in Section V, whereby we summarize our main conclusions. An appendix is also included, in which the field equations for the homogeneous and isotropic flat universe are provided.

Before proceeding to the next section, we would like to note that  in EsGB gravity GWs produced by astrophysical sources have been studied  by various authors \cite{Daniel24,Nojiri24a,Nojiri24b,Evstafyeva23,Wong22,Lyu:2022gdr,Shiralilou22,Carson20,JLB16,Yagi:2012gp}. However, to our best knowledge, this is the first time in the literature to work out the condition (\ref{CD1}) for the theory to be free from ghosts and Laplacian instability, and obtain the severest observational constraint (\ref{eq2.4uu}) for $c_T = 1$ in all epochs of the evolution of the universe. It is also the first time to obtain the explicit expressions of the amplitudes of the spin-0 and spin-2 gravitons along their trajectories in the homogeneous and isotropic flat universe.

 In this paper we shall adopt the conventions, in which the signature of the metric is ($-, +, +, +$), while the Christoffel symbols, Riemann and Ricci tensors, as well as the Ricci scalar,  are defined, respectively, by
\begin{eqnarray}
\label{eq1.3}
&& \Gamma^\alpha_{\mu\nu} \equiv \frac{1}{2}g^{\alpha\beta}\left(g_{\beta\nu, \mu} + g_{\beta\mu, \nu} - g_{\mu\nu, \beta}\right)\,,\nonumber\\
&& \left(D_{\alpha}D_{\beta} - D_{\beta}D_{\alpha}\right) X^{\mu} = {R^{\mu}}_{\nu\alpha\beta}X^{\nu}\,, \nonumber\\
&& R_{\mu\nu} \equiv R^{\alpha}_{\;\;\mu\alpha\nu}, \quad R \equiv g^{\mu\nu} R_{\mu\nu}\,,
\end{eqnarray}
where $D_{\alpha}$ denotes the covariant derivative with respect to  metric $g_{\mu\nu}$, and $g_{\mu\nu,\lambda} \equiv \partial g_{\mu\nu}/\partial x^{\lambda}$, and
\begin{eqnarray}
R^{\alpha}_{\;\;\mu\nu\lambda} &=& \Gamma^{\alpha}_{\mu\lambda,\nu}- \Gamma^{\alpha}_{\mu\nu,\lambda}
+ \Gamma^{\alpha}_{\beta\nu}\Gamma^{\beta}_{\mu\lambda} - \Gamma^{\alpha}_{\beta\lambda}\Gamma^{\beta}_{\mu\nu}\,.
\end{eqnarray}

\section{Einstein scalar Gauss Bonnet Gravity and Observational Constraints}
\renewcommand{\theequation}{2.\arabic{equation}} \setcounter{equation}{0}

In this section, we shall first give a very brief introduction to EsGB gravity, and then find the conditions for which the theory is absent of ghosts and Laplacian instability \cite{Tsujikawa:2022aar}, and meanwhile it is consistent with all observations carried out so far, as in the literature there still exist some mutually contradicting statements.

 \subsection{EsGB Gravity}

The   EsGB gravity is described by the action \cite{Hussain24,Minamitsuji24,Zhang20a}
\begin{eqnarray}
S_{\text{EsGB}} &=& \frac{1}{2\kappa}\int dx^4\sqrt{-g}\Big[R-2\Lambda  +\alpha f(\phi){\cal G}\nonumber\\
&& ~~~~~~~~  +{\cal L}_{\phi}\left(\phi\right) + 2\kappa  {\cal L}_{m}\left(g_{\mu\nu}; \psi\right)\Big]\,,
\end{eqnarray}
where $\kappa  \equiv 8\pi G/c^4$, $G$ is the Newtonian constant, $c$ the speed of light, and $g \;[\equiv \text{det}(g_{\mu\nu})]$ is the determinant of $g_{\mu\nu}$.
The scalar field $\phi$ is non-minimally coupled to the Gauss-Bonnet term ${\cal G}$
\begin{eqnarray}
{\cal G}&\equiv&R^2 -4R_{\mu\nu}R^{\mu\nu} +R_{\mu\nu\rho\sigma}R^{\mu\nu\rho\sigma}\,,
\end{eqnarray}
through the arbitrary function $f(\phi)$ with a coupling constant $\alpha$, and
$R$ and $\Lambda$ denote  the Ricci scalar and   the cosmological constant,  respectively. ${\cal L}_{\phi}$ is the Lagrangian density for  the scalar field with
\begin{eqnarray}
{\cal L}_\phi&=&-\frac{1}{2}\left(\partial^\mu\phi\right)\left(\partial_\mu\phi\right)-V(\phi)\,, 
\end{eqnarray}
where $V(\phi)$ is the potential of the scalar field and $\partial^{\mu}=g^{\mu\nu}\partial_{\nu}$. 
The  Lagrangian density  ${\cal L}_{m}$ represents
both the matter fields that produce our inhomogeneous  universe and the astrophysical sources  that produce GWs, such as binary systems of compact astrophysical objects,
 to be studied in this paper. To avoid the fifth force, we assume that only the metric $g_{\mu\nu}$ is minimally coupled with matter, collectively denoted by $\psi$.

From the above action, one can derive the equations of motion for both $g_{\mu\nu}$ and $\phi$,
\begin{eqnarray}
&& G_{\mu\nu} + \Lambda g_{\mu\nu}=\alpha T^{GB}_{\mu\nu}+T^\phi_{\mu\nu} +\kappa T^m_{\mu\nu}\,, ~~~\\
&& D^2\phi - V_{,\phi} = - \alpha f_{,\phi}{\cal G}\,,
\end{eqnarray}
where $f_{,\phi} \equiv df(\phi)/d\phi$, $G_{\mu\nu} \left(\equiv R_{\mu\nu}-\frac{1}{2}g_{\mu\nu}R\right)$ denotes the Einstein tensor, $D^2 \equiv g^{\mu\nu}D_{\mu}D_{\nu}$, $D_{\mu}$ is the covariant derivative with respect to  metric $g_{\mu\nu}$ as in (\ref{eq1.3}), $V_{,\phi} \equiv dV(\phi)/d\phi$, and 
\begin{eqnarray}
T^{GB}_{\mu\nu}&\equiv&2\left(D_\mu D_\nu f\right)R-2g_{\mu\nu}\left(D_\rho D^\rho f\right)R \nonumber\\
		&&-4\left(D^\rho D_\nu f\right)R_{\mu\rho}-4\left(D^\rho D_\mu f\right)R_{\nu\rho}\nonumber\\
		&&+4\left(D^\rho D_\rho f\right)R_{\mu\nu}+4g_{\mu\nu}\left(D^\rho D^\sigma f\right)R_{\rho\sigma}\nonumber\\
		&&-4\left(D^\rho D^\sigma f\right)R_{\mu\rho\nu\sigma}\,,\nonumber\\
T^\phi_{\mu\nu}&\equiv&\frac{1}{2}\left(D_\mu\phi\right) D_\nu\phi-\frac{1}{4} g_{\mu\nu}\Big[\left(D^\alpha\phi\right)D_\alpha\phi + 2V(\phi)\Big],\nonumber\\
T^m_{\mu\nu} &\equiv& - \frac{2}{\sqrt{-g}}\frac{\delta\left(\sqrt{-g}\; {\cal{L}}_{m}\right)}{\delta g^{\mu\nu}}\,.
\end{eqnarray}
Note that the matter field does not couple with the scalar field $\phi$ directly, so we have the conservation law,
\begin{eqnarray}
D^{\nu} T^m_{\mu\nu} = 0\,.
\end{eqnarray}
Then, the contracted Bianchi identities $D^{\nu}G_{\mu\nu} = 0$     
lead to
\begin{eqnarray}
D^{\nu}\left(T^\phi_{\mu\nu} +\alpha T^{GB}_{\mu\nu}\right) = 0\,.
\end{eqnarray}
It should be also noted that in general $D^{\nu}T^\phi_{\mu\nu} \not= 0$, due to the coupling  between the scalar field and the Gauss-Bonnet term ${\cal{G}}$.

\subsection{Stability  Conditions and Observational Constraints}

To consider the conditions for EsGB gravity to be physically viable,
let us consider the flat FLRW universe
\begin{eqnarray}
\label{eq2.4cc}
ds^2 = - dt^2 + a^2(t)\left(dx^2 + dy^2 + dz^2\right)\,,  
\end{eqnarray}
where $t$ denotes the cosmic time of the universe and $x^{\mu} = \left(t, x, y, z\right)$. Then, the corresponding field equations of EsGB gravity are presented in Appendix \ref{AppendixA}. In this background,  the kinetic coefficients $q_T$ and $q_S$ of the tensor and scalar modes as well as their speeds $c_T$ and $c_S$ are given respectively by  \cite{Tsujikawa:2022aar} \footnote{It can be shown that the relations between $(\tilde{q}_t, \tilde{q}_s)$ used in  \cite{Tsujikawa:2022aar} and the ones $(q_T, q_S)$ used in this paper are given by $\tilde{q}_t = q_T/\kappa,\; \tilde{q}_s = q_S/\kappa^2$.}
\begin{eqnarray}
\label{eq2.4dd}
q_T &=&  1 + 4\alpha H \dot{f},  \quad q_S  = q_T + 48\alpha^2 H^4 f_{,\phi}^2\,, \nonumber\\
c_T^2 &=&  1 + \frac{\alpha\left(\ddot{f} - H \dot{f}\right)}{q_T}\,, \nonumber\\
c_S^2 &=& \frac{1}{q_S}\left[q_T - 16\left(2 + c_T^2 + 6 w_{\text{eff}}\right)\alpha^2 H^4 f_{,\phi}^2\right]\,, ~~~~~~~~~
\end{eqnarray}
where $\dot{f} \equiv df/dt = f_{,\phi}\dot{\phi}$, $\dot{\phi} = d\phi/dt$, and 
\begin{eqnarray}
   w_{\text{eff}} \equiv - 1 + \frac{2}{3} \epsilon_H,  \quad   \epsilon_H \equiv - \frac{\dot H}{H^2}\,. ~~~~~~~~
\end{eqnarray}
The conditions to be free from ghosts and Laplacian instability require 
\begin{eqnarray}
\label{eq2.4ff}
q_T, \; q_S, \;  c_T^2, \; c_S^2 > 0\,. 
\end{eqnarray}
To satisfy these conditions, the following conditions are usually requested  \cite{Tsujikawa:2022aar}
\begin{eqnarray}
\label{eq2.4gg}
\left\{\left|\alpha f_{,\phi}H\dot\phi\right|, \left|\alpha f_{,\phi}\ddot\phi\right|, \left|\alpha f_{,\phi\phi}\dot\phi^2\right|\right\} \ll 1\,,
\end{eqnarray}
where $H \equiv \dot{a}/a$.

In addition, the constraint from the observation of the gravitational wave event GW170817 \cite{LIGOScientific:2017zic}
\begin{eqnarray}
\label{CD}
-\; 3\times10^{-15} \lesssim c_T - 1 \lesssim 7\times 10^{-16}\,,
\end{eqnarray}
requires
\begin{eqnarray}
\label{eq2.4hh}
\left|\alpha f_{,\phi}\ddot\phi + \alpha f_{,\phi\phi}\dot\phi^2 - \alpha f_{,\phi}H\dot\phi\right| \lesssim  10^{-15}\,.  
\end{eqnarray}
It can be shown that with the above conditions we have
$c_T \simeq 1$. 
In the rest of this section, we shall set  
\begin{eqnarray}
\label{eq2.4hhaa}
c_T = 1\,.
\end{eqnarray}
With the above condition, we shall show that Conditions (\ref{eq2.4gg}) and (\ref{eq2.4hh}) are satisfied, provided that the condition (\ref{CD1}) to be given below  holds.
 To prove the above claim, let us first note that the condition (\ref{eq2.4hhaa}) implies 
\begin{eqnarray}
\alpha\dot{f} = \hat\alpha a\,,
\end{eqnarray}
as can be seen from Eq.(\ref{eq2.4dd}), where $\hat\alpha$ is a constant. 
Then, we find 
\begin{eqnarray}
\label{eq2.4kk}
q_T = 1 + 4\hat\alpha\dot{a}\,.
\end{eqnarray} 
Therefore, in an expanding universe, the condition $q_T > 0$ holds, provided that  
\begin{equation}
\label{CD1}
\hat\alpha \geq  0\,.
\end{equation}
This condition also leads to $q_S > 0$, as can be seen from Eq.(\ref{eq2.4dd}). Therefore, we conclude that {\em the condition (\ref{CD1}) guarantees that 
the EsGB gravity is free from ghost instability}, when (\ref{eq2.4hhaa}) is imposed.

On the other hand, since now $c_T = 1$, the Laplacian instability is also absent for the spin-2 gravitons. To show that this is also the case for the spin-0 gravitons, 
let us first note that $f'(\phi) = \dot{f}/\dot\phi$, so that $c_S^2$ can be cast in the form
\begin{eqnarray}
\label{eq2.4ee}
c_S^2 &=& \frac{{\cal{D}}}{q_S\dot\phi^2}\,,\nonumber\\
   {\cal{D}} &\equiv& q_T \dot\phi^2 + 48\hat\alpha^2 a^2H^4\left(1 - \frac{4}{3}\epsilon_H\right)\,. ~~~~~~~~
\end{eqnarray}
Then,   in the inflationary or current (dark-energy dominated) epoch, we have $\left|\epsilon_H\right| \ll 1$, so the function ${\cal{D}}$ defined above   is
always positive. As a result, $c_S^2 \ge 0$.

In the radiation-dominated epoch, we have
\begin{eqnarray}
a(t) = a_{\text{eq}}\left(\frac{t}{t_{\text{eq}}}\right)^{1/2},\quad \rho \simeq \rho_R = \frac{\rho_R^0}{a^4}\,,
\end{eqnarray}
where $a_{\text{eq}} \equiv a(t_{\text{eq}})$, $t_{\text{eq}}$ denotes the time when the radiation energy density was equal to the matter energy density, and
$\rho_R^0$ is a constant. 
Then, we find
\begin{eqnarray}
\label{eq2.4mm}
q_T(t) = 1 + \left(\frac{t_{Rc}}{t}\right)^{1/2}, \quad t_{Rc} \equiv \left(\frac{2\hat\alpha a_{\text{eq}}}{t_{\text{eq}}^{1/2}}\right)^{2}\,.  
\end{eqnarray}
  Combining the above expressions with the field equation (\ref{eq3.15}) presented in Appendix \ref{AppendixA}, we find that
\begin{eqnarray}
\label{eq2.4nn}
\dot\phi^2(t) &=& \frac{2}{t^2}\left\{\left(1 - \epsilon_R\right) + \left(\frac{t_{Rc}}{t}\right)^{1/2}\right\}\,,
\end{eqnarray}
for which Eq.(\ref{eq2.4ee}) yields
\begin{eqnarray}
\label{eq2.4nnD}
{\cal{D}}(t) &=& \frac{1}{\kappa^2 t^2\dot\phi^2}\left\{\left(1 - \epsilon_R\right) + \left(2 - \epsilon_R\right)\left(\frac{t_{Rc}}{t}\right)^{1/2}\right.\nonumber\\
&& ~~~~~~~~~~~~~~ \left. + \frac{3}{8} \left(\frac{t_{Rc}}{t}\right)\right\}\,, 
\end{eqnarray}
where 
\begin{eqnarray}
\label{eq2.4oo}
\epsilon_R \equiv \frac{64\kappa  \hat\alpha^4 \rho_R^0}{3t_{Rc}^2} = 4H_0^2\Omega^{(0)}_R \left(\frac{t_{\text{eq}}}{a_{\text{eq}}^2}\right)^2\,. 
\end{eqnarray}
Then, for \cite{Liddle2015}
\begin{eqnarray}
a_{\text{eq}} &=& \frac{1}{2.4 \times 10^4 \Omega_0 h^2}, \quad t_{\text{eq}} = \frac{3.4\times 10^3}{\left(\Omega_0 h^2\right)^{3/2}}\; (\text{yrs})\,, \nonumber\\
H_0 &=& \frac{h}{9.77 \times 10^9} \;(\text{yrs})^{-1}\,, 
\end{eqnarray}
we obtain
\begin{eqnarray}
\epsilon_R \simeq 1.614 \times 10^5\left(\Omega_0 h^2\right) \left(\Omega^{(0)}_R h^2\right) \simeq 0.586\,,   
\end{eqnarray}
for $\Omega^{(0)}_R h^2 \simeq 2.47\times 10^{-5},\; \Omega_0 h^2 \simeq 0.147$ \cite{Liddle2015}. Thus, from Eq.(\ref{eq2.4nn}) we find that the condition ${\cal{D}}(t) > 0$
in the radiation-dominated epoch holds, so does the condition $c_S^2 > 0$.
 
 In the matter-dominated epoch, we have
\begin{eqnarray}
a(t) = a_{D}\left(\frac{t}{t_D}\right)^{2/3},\quad \rho \simeq \rho_M = \frac{\rho_M^0}{a^3}\,,
\end{eqnarray}
where $a_{D} \equiv a(t_{D})$, $t_{D}$ denotes the transition time from the matter-dominated epoch to the DE-dominated one, and $\rho_M^0$ is a constant. Then, we find 
\begin{eqnarray}
\label{eq2.4ss}
q_T(t) = 1 + \left(\frac{t_{Dc}}{t}\right)^{1/3}, \quad t_{Dc} \equiv \left(\frac{8\hat\alpha a_{D}}{3t_{D}^{2/3}}\right)^{3}\,. 
\end{eqnarray}
Hence, we obtain
\begin{eqnarray}
\label{eq2.4tt}
\dot\phi^2(t) &=& \frac{8}{3t^2}\left\{\left(1 - \epsilon_D\right) + \left(\frac{t_{Dc}}{t}\right)^{1/3}\right\}\,,\nonumber\\
{\cal{D}}(t) &=& \frac{4}{3\kappa^2 t^2\dot\phi^2}\left\{\left(1 - \epsilon_D\right) + \left(2 - \epsilon_D\right)\left(\frac{t_{Dc}}{t}\right)^{1/3}\right.\nonumber\\
&& ~~~~~~~~~~~~~~ \left. + \frac{1}{2} \left(\frac{t_{Dc}}{t}\right)^{2/3}\right\}\,, 
\end{eqnarray}
where 
\begin{eqnarray}
\label{eq2.4vv}
\epsilon_D &\equiv& \frac{3\kappa  \rho_M^0 t_D^2}{4 a_D^3}  \nonumber\\
&\simeq& 2.357 \times 10^{-20}\left(\Omega^{(0)}_{\text{DE}} h^2\right) \frac{t_D^2}{\left(\text{yrs}\right)^2}\,.
\end{eqnarray}
Thus, for   $\Omega^{(0)}_{\text{DE}} h^2 \simeq 0.343, \; t_D \simeq 9\times 10^9$ yrs \cite{Liddle2015}, we have $\epsilon_D \simeq 0.65484$.  Then, Eq.(\ref{eq2.4tt}) tells us that the condition ${\cal{D}}(t) > 0$ also holds in the matter-dominated epoch. 

Therefore, we conclude that {\em when $c_T = 1$ the conditions (\ref{eq2.4ff}) hold in all epochs of the evolution of the Universe, provided that the condition (\ref{CD1}) is true} \footnote{It must be noted that the observation of GW170817 happened in the very low redshift. In the high redshift epoch,  the constraint (\ref{CD}) may not be applicable. Then, it is not necessary to set $c_T = 1$, meaning that the parameter space where (\ref{eq2.4ff}) holds is even larger.}.

In addition, solar system tests lead to the following constraint \cite{Amendola:2007ni}
\begin{eqnarray}
\left|\alpha f_{,\phi}(\phi_0)\right| \lesssim 1.6 \times 10^{14}\; (\text{km})^2\,, 
\end{eqnarray}
but the observations of low-mass x-ray binary and gravitational waves imposed much severer  constraint \cite{Yagi:2012gp,Lyu:2022gdr}
\begin{eqnarray}
\label{eq2.4ww}
\left|\alpha f_{,\phi}(\phi_0)\right| \lesssim 1.18\; (\text{km})^2\,,
\end{eqnarray}
where $\phi_0$ is the current value of the scalar field $\phi$, and $\alpha f_{,\phi}(\phi_0) = \alpha \dot{f}(t_0)/\dot\phi(t_0) = \hat{\alpha} a_0 / \dot{\phi}(t_0)$. Considering the field equation (\ref{eq3.14}) we find that $\left|\dot\phi(t_0)\right| \simeq {\cal{O}}\left(H_0\right)$. So, Eq.(\ref{eq2.4ww}) implies
\begin{eqnarray}
\label{eq2.4uu}
\left|\hat\alpha\right| \lesssim 1.18\times H_0\; (\text{km})^2 \simeq 8.97 \times 10^{-24} \; (\text{km})\,.
\end{eqnarray}
This represents the severest observational  constraint found so far for EsGB gravity.

\section{Gravitational Waves Propagating in the Homogeneous Universe}
 \label{SecIII}
 \renewcommand{\theequation}{3.\arabic{equation}} \setcounter{equation}{0}

In this section, we consider  GWs propagating in the homogeneous and isotropic background for a long time, before reaching to the Earth for us to detect. The flat FLRW universe is described by the metric (\ref{eq2.4cc}). Adding the perturbation $h_{\mu\nu}$ and in terms of the conformal time $\eta \equiv \int dt/a(t)$, it reads
\begin{equation}
 g_{\mu\nu} = \gamma_{\mu\nu} + h_{\mu\nu}\,, 
\end{equation}
where
\begin{eqnarray}
\label{CosmologicalMetric}
 \gamma_{\mu\nu} dx^{\mu} dx^{\nu} = a^2(\eta)\left(-d\eta^2+ dx^2+dy^2+dz^2\right)\,.
\end{eqnarray}
The basic field equations for the flat FLRW background are presented in Appendix \ref{AppendixA}.

\subsection{Scalar and Tensor Decomposition of Gravitational Waves} 

Let us consider GWs produced by remote astrophysical sources in the flat FLRW Universe, and then propagating long distances before arriving to us. In such a background,  the GWs can be uniquely decomposed into scalar and tensor modes \cite{DB09,KAM23} \footnote{Similarly to GR, the vector perturbations in EsGB gravity can be set to zero identically \cite{Cartier:2001is}.}
\begin{eqnarray}
    h_{\mu\nu}&=&a^2\left[-2\omega\delta^0_{\mu} \delta^0_{\nu} + (\delta^0_{\mu} \delta^i_{\nu} + \delta^i_{\mu} \delta^0_{\nu})\partial_i S + 2\varpi \delta^i_{\mu} \delta^j_{\nu} \right. \nonumber\\
    & & \left. ~~~~~ + 2\partial_i \partial_j E \delta^i_{\mu} \delta^j_{\nu} + H_{ij} \delta^i_{\mu} \delta^j_{\nu}\right]\,,
\end{eqnarray}
where $x^{\mu} \equiv (\eta, x^i)$, and 
\begin{eqnarray}
&& 
\delta^{ij}H_{ij} = 0, \quad \delta^{ij}\partial_i H_{jk} = 0\,.
\end{eqnarray}
 The advantage of the decomposition is that the linearized field equations will be decoupled between the scalar and tensor modes \cite{DB09,KAM23}. The tensor mode $H_{ij}$ is gauge invariant, while for the scalar modes, we can construct the following gauge invariants
\begin{eqnarray}
\label{2.6cc}
&& \Phi_B \equiv \omega + {\cal{H}}\left(S - E'\right) + \left(S - E'\right)'\,,    \nonumber\\
&& \Psi_B \equiv \varpi +  {\cal{H}}\left(S - E'\right)\,,\nonumber\\
&& \Phi \equiv  \varpi - \frac{\cal{H}}{\bar\phi'} \varphi\,. 
\end{eqnarray}
It is clear that $\Phi_B$ and $\Psi_B$ are directly related to the metric perturbations, while $\Phi$ to the scalar field perturbation. 

To study the propagation of GWs over the homogeneous and isotropic universe further, let us consider the tensor and scalar modes separately. Furthermore, it is convenient to Fourier transform all perturbation variables w.r.t. the comoving spatial coordinates $x^i$. Thanks to the homogeneity of the background, different Fourier modes are decoupled from each other at the level of the linear perturbations.

\subsection {Tensor Modes}

In the tensor sector we have only $H_{ij}$ and we denote its Fourier transformation as $\tilde{H}_{ij}$. Then, we find that $\tilde{H}_{ij}$ satisfies the equation  \cite{HN05}
\begin{eqnarray}
\label{eq3.19aa}
 \frac{1}{a^2q_T}\partial_{\eta}\left(a^2q_T \partial_{\eta}\tilde{H}_{ij}\right) + c_T^2\frac{\delta^{ij}k_ik_j}{\epsilon^2} \tilde{H}_{ij} = 0\,,
\end{eqnarray}
where $q_T$ and $c_T$ are given in Eq.(\ref{eq2.4dd}), $k_i/\epsilon$ is the comoving wave vector and $\epsilon$ is a small bookkeeping parameter introduced in order implement the geometrical optics approximation

We then adopt the WKB-type ansatz
\begin{equation}
 \tilde{H}_{ij} = {\cal{A}}_T e_{ij} e^{i \Theta_T/\epsilon}\,,
\end{equation}
where $\Theta_T/\epsilon$ and $e_{ij}$ denote respectively the phase of the tensor mode and a constant polarization tensor satisfying 
\begin{eqnarray}
 \delta^{ij}e_{ij} = 0\,, \quad \delta^{ij}k_ie_{jk} = 0\,, \quad \delta^{ik}\delta^{jl}e_{ij}e_{kl} = 1\,.
\end{eqnarray}
We then have
\begin{eqnarray}
 \partial_{\eta}\tilde{H}_{ij} &=& \left({\cal{A}}'_T - \frac{i\omega_T}{\epsilon}{\cal{A}}_T\right)e_{ij} e^{i\Theta_T/\epsilon}\,, \nonumber\\
\partial^2_{\eta}\tilde{H}_{ij} &=& \Bigg[{\cal{A}}''_T - \frac{i}{\epsilon}\left(2\omega_T{\cal{A}}'_T + \omega'_T{\cal{A}}_T\right) - \frac{\omega_T^2}{\epsilon^2}{\cal{A}}_T\Bigg] \nonumber\\
&& \times e_{ij} e^{i\Theta_T/\epsilon}\,,
\end{eqnarray}
where $\omega_T = -\partial_{\eta}\Theta_T$. Thus, to the leading order $\epsilon^{-2}$, from Eq.(\ref{eq3.19aa}) we find $\left(\omega_T^2-c_T^2\delta^{ij}k_ik_j\right) {\cal{A}}_T = 0$. Since ${\cal{A}}_T \not= 0$, we must have
\begin{eqnarray}
\label{eq3.19cc}
\epsilon^{-2}: \; \omega_T^2-c_T^2\delta^{ij}k_ik_j = 0\,.
\end{eqnarray}
On the other hand, to the order of $\epsilon^{-1}$, Eq.(\ref{eq3.19aa}) yields
\begin{eqnarray}
\label{eq3.19dd}
\epsilon^{-1}: \; 2\omega_T{\cal{A}}'_T + \omega'_T{\cal{A}}_T + \omega_T{\cal{A}}_T\partial_{\eta}\ln (a^2q_T) = 0\,. 
\end{eqnarray}

To study further the propagation of the spin-2 gravitons in EsGB gravity, let us consider  the case $c_T =  1$.  When $c_T = 1$, Eq.(\ref{eq3.19cc}) reduces to 
\begin{equation}
    \gamma^{\mu\nu}k_{\mu}k_{\nu} = 0\,,
\end{equation}
where $k_{\mu} \equiv (-\omega_T, k_i) = \partial_{\mu}(-\Theta_T + k_ix^i)$. Hence, similarly to the GR case \cite{Fier21}, {\it the spin-2 gravitons in EsGB gravity with $c_T=1$ are also moving along null geodesics}. This is not surprise, as we already assumed  $c_T = 1$. Introducing the null trajectories defined by
\begin{eqnarray}
\frac{dx^{\mu}(\lambda)}{d\lambda} = k^{\mu}\,,
\end{eqnarray}
where $k^{\mu}\equiv \gamma^{\mu\nu}k_{\nu} = (\omega_T/a^2, \delta^{ij}k_j/a^2)$ and $\lambda$ denotes the affine parameter along the null geodesics, we find that
\begin{eqnarray}
\frac{Dk_{\mu}}{D\lambda} = k^{\alpha}\nabla_{\alpha}k_{\mu} = \frac{1}{2}\partial_{\mu}(k^{\alpha}k_{\alpha}) = 0\,.
\end{eqnarray}
Here, we have used the fact that $\nabla_{\alpha}k_{\mu} = \nabla_{\alpha}\nabla_{\mu}(-\Theta_T+k_ix^i) = \nabla_{\mu}k_{\alpha}$.

With $c_T=1$, the propagation equation in EsGB theory to the order of $\epsilon^{-1}$, given by (\ref{eq3.19dd}), can be written as
\begin{eqnarray}\label{tensor-amplitude-EsGB}
\partial_{\eta}\left( a {\cal{A}}_T\sqrt{q_T}\right) = 0\,.
\end{eqnarray}
On the other hand, in GR we have
\begin{eqnarray}\label{tensor-amplitude-GR}
\partial_{\eta}\left( a {\cal{A}}_T^{(GR)}\right) = 0\,.
\end{eqnarray}
 Thus, integrating Eq.(\ref{tensor-amplitude-EsGB}) and (\ref{tensor-amplitude-GR}), we obtain  
 \begin{equation}
{\cal{A}}_T =  {\cal{A}}_T^{\text{(GR)}} \left(\frac{q_{te}}{q_T(t)}\right)^{1/2}\,, \quad 
{\cal{A}}_T^{\text{(GR)}} \equiv {\cal{A}}_e\left(\frac{{\cal{R}}_e}{{\cal{R}}(t)}\right)\,,  
\end{equation}
where    
${\cal{R}}(t) \left(\equiv r_e a(t)\right)$  denotes the physical distance between the source and observer measured in the flat FLRW spacetime, and $r_e \left(\equiv \sqrt{x_e^2 + y_e^2 + z_e^2}\right)$ is the comoving distance between the observer at $x^i=0$ and the source at $x^i=(x_e, y_e, z_e)$. The constants ${\cal{R}}_e \;(\equiv r_e a(t_e))$,
$q_{Te}$ and ${\cal{A}}_e$  are the values of  ${\cal{R}}$, $q_T$ and ${\cal{A}}_T$ (or ${\cal{A}}_T^{\text{(GR)}}$) at the emission time  $t_e$.  
On the other hand, from  Eq.(\ref{eq2.4kk}) we find that
 \begin{eqnarray}
 q_T(t) = 1 + 4\hat\alpha \dot{a} = \begin{cases}
  1 + \left(\frac{t_{Rc}}{t}\right)^{1/2}, & \text{RD}\,,\cr 
   1 + \left(\frac{t_{Dc}}{t}\right)^{1/3}, & \text{MD}\,,\cr
    1 + 4\hat\alpha H_0 e^{H_0(t - t_0)}, & \text{$\Lambda$D}\,,\cr 
 \end{cases}
\end{eqnarray}
where $t_{Rc}$ and $t_{Dc}$ are given in Eqs.(\ref{eq2.4mm}) and (\ref{eq2.4ss}), $H_0$ and $t_0$ are respectively the current Hubble parameter and time. It is interesting to note that $q_T(t)$ is decreasing as $t$ increases in both of the radiation and matter dominated epochs, while it is increasing during the DE dominated epoch. 

To see the specific effects of the EsGB theory on the propagation of GWs, let us first recall the theoretical and observational constraints on $\hat\alpha$, given  by Eqs.(\ref{CD1}) and (\ref{eq2.4uu}), respectively. We then show that the deviation from GR is negligible. For this purpose we consider a situation which is astrophysically/cosmologically unrealistic but which gives a rather conservative upper bound on the deviation from GR. We suppose that a GW was emitted at the beginning of the radiation epoch, e.g. at the time of reheating after inflation $t=t_{\text{reh}}$, and estimate the deviation from GR accumulated till the end of radiation dominated epoch, i.e. the matter-radiation equality $t=t_{\text{eq}}$. We then find that 
\begin{eqnarray}
\frac{q_{\text{Te}}(t_{\text{reh}})}{q_T(t_{\text{eq}})}
= \frac{t^{1/2}_{\text{reh}} + t^{1/2}_{\text{Rc}}}{t^{1/2}_{\text{eq}} + t^{1/2}_{\text{Rc}}}\left(\frac{t_{\text{eq}}}{t_{\text{reh}}}\right)^{1/2}\,,
\end{eqnarray}
where
\begin{eqnarray}
t_{\text{Rc}} &=& \left(\frac{2\hat\alpha a_{\text{eq}}}{t^{1/2}_{\text{eq}}}\right)^{2} \simeq 5.3\times 10^{-12}\; \left(\frac{\hat\alpha^2}{\text{yrs}}\right)\nonumber\\
&\lesssim& 4.78 \times 10^{-84} \; (\text{yrs}) \ll t_{\text{reh}}  \ll t_{\text{eq}}\,.  
\end{eqnarray}
Therefore, we find that 
\begin{eqnarray}
\frac{q_{\text{Te}}(t_{\text{reh}})}{q_T(t_{\text{eq}})} - 1
&=& \frac{t^{1/2}_{\text{reh}} + t^{1/2}_{\text{Rc}}}{t^{1/2}_{\text{eq}} + t^{1/2}_{\text{Rc}}}\left(\frac{t_{\text{eq}}}{t_{\text{reh}}}\right)^{1/2} - 1 \nonumber\\
&\simeq& {\cal{O}}\left(\frac{t^{1/2}_{\text{Rc}}}{t^{1/2}_{\text{reh}}}\right)
\lesssim {\cal{O}}\left(10^{-22}\right)\,,
\end{eqnarray}
that is, the modifications of the propagation of GWs in the radiation-dominated epoch are negligible in EsGB theory in comparing them with those given  in GR. Here, we have supposed the lower bound on the time of reheating as $t_{\text{reh}}\gtrsim 3.2\times 10^{-40}$ (yrs).

On the other hand, we also have
\begin{eqnarray}
t_{\text{Dc}} &=& \left(\frac{8\hat\alpha a_{D}}{3t^{2/3}_{D}}\right)^{3} \simeq 6.96\times 10^{-20}\; \left(\frac{\hat\alpha^3}{(\text{yrs})^2}\right)\nonumber\\
&\lesssim& 5.96 \times 10^{-128} \; (\text{yrs}) \ll t_{\text{eq}}  \ll t_{\text{D}}\,,\nonumber\\
\hat\alpha H_0 &\lesssim& 6.8 \times 10^{-47}. 
\end{eqnarray}
Again, we consider a situation that is unrealistic from astrophysical/cosmological viewpoints but that gives a rather conservative upper bound on the deviation from GR, in order to show that the deviation is negligible also in the matter dominated epoch. To maximize the deviation from GR, we suppose that a GW is emitted at the beginning of the matter dominated epoch, i.e. the matter-radiation equality $t=t_{\text{eq}}$, and then observed at the end of the matter dominated epoch, i.e. the transition time from the matter-dominated epoch to the DE-dominated one $t=t_{D}$. Then, we find that the maximum deviation from GR during the matter dominated epoch is 
\begin{eqnarray}
\frac{q_{\text{Te}}(t_{\text{eq}})}{q_T(t_{D})} - 1
&=& \frac{t^{1/3}_{\text{eq}} + t^{1/3}_{\text{Dc}}}{t^{1/3}_{D} + t^{1/3}_{\text{Dc}}}\left(\frac{t_{\text{D}}}{t_{\text{eq}}}\right)^{1/3} - 1 \nonumber\\
&\simeq& {\cal{O}}\left(\frac{t^{1/3}_{\text{Dc}}}{t^{1/3}_{\text{eq}}}\right)
\lesssim {\cal{O}}\left(10^{-44}\right)\,.
\end{eqnarray}

Similarly, the maximum deviation from GR during the DE dominated epoch is 
\begin{eqnarray}
\frac{q_{\text{Te}}(t_{\text{D}})}{q_T(t_{0})} - 1
&=& \frac{1 + 4\hat\alpha H_0 e^{-H_0(t_0 - t_D)}}{1 + 4\hat\alpha H_0} - 1 \nonumber\\
&\simeq& {\cal{O}}\left(\hat\alpha H_0\right) \lesssim {\cal{O}}\left(10^{-46}\right)\,.
\end{eqnarray}
Thus,   we conclude that {\em in the whole history of the evolution of the homogeneous and isotropic Universe from the onset of the radiation dominated epoch, the deviations of the propagation of GWs between EsGB gravity and GR are negligible}.

\subsection {Scalar Modes}

In the scalar sector, there are three gauge invariant variables (\ref{2.6cc}), that are constructed from five perturbation variables by eliminating two gauge degrees of freedom. Needless to say, working with the gauge-invariant variables is equivalent to working with variables remaining after gauge fixing. For example, one can set 
\begin{eqnarray}
 S = E = 0\,, \quad (\mbox{Newtonian gauge})\,,
\end{eqnarray}
for which  Eq.(\ref{2.6cc}) yields
\begin{equation}
 \Phi_B = \omega\,, \quad \Psi_B = \varpi\,, \quad \Phi = \varpi - \frac{\cal{H}}{\bar\phi'} \varphi\,.
\end{equation}
There are five equations of motion corresponding to variations of the action w.r.t. the five perturbation variables but two among them are dependent of the others due to two components of the Bianchi identity in the scalar sector, resulting in three independent equations of motion. By using two among three independent equation of motion, $\Phi_B$ and $\Phi$ can be expressed in terms of $\Psi_B$ and its derivative. The remaining independent equation of motion leads to the field equation for $ \Psi_B$ of the form  \cite{HN05}
\begin{eqnarray}
\label{eq4.52}
\frac{I}{JK}\left[\frac{J^2}{a I}\left(\frac{aK}{J}\Psi_B\right)_{,t}\right]_{,t} = \frac{c_S^2}{a^2}\delta^{ij}\partial_i\partial_j\Psi_B\,,
\end{eqnarray}
where
\begin{eqnarray}
I &\equiv& \frac{\cal{D}}{2\kappa q_T}, \quad J \equiv \frac{H}{q_T}\left(1 + 6 \hat\alpha aH\right)\,,\nonumber\\
K&\equiv& \frac{q_T}{\kappa}, \quad \Delta \equiv q_T\dot\phi^2 + 48 \hat\alpha^2 a^2H^4,\nonumber\\
c_S^2 &\equiv& 1 - \alpha_S, \quad \alpha_S \equiv  \frac{64 \hat\alpha^2 }{\Delta}a^2H^4\epsilon_H\,.
\end{eqnarray}
Here, ${\cal{D}}$ and $q_T$ are given by Eqs.(\ref{eq2.4ee}) and (\ref{eq2.4kk}), and $\dot\phi^2$ is given by Eq.(\ref{eq3.15}), from which we find 
\begin{eqnarray}
\dot\phi^2 = - 2\left[2q_T\dot{H} + \kappa\left(\rho + p\right)\right]\,.
\end{eqnarray}
Recall that $\rho$ and $p$ denote the energy density and pressure of the cosmic fluid. 
Setting 
\begin{eqnarray}
\label{eq4.56}
\hat{\Psi}_B \equiv \frac{aK}{J}\Psi_B, \quad \hat{q}_s \equiv \frac{J^2}{a^4 I}\,,
\end{eqnarray}
we find that Eq.(\ref{eq4.52}) can be cast in the form 
\begin{eqnarray}
\label{eq4.57}
\frac{1}{a^2 \hat{q}_s}\partial_{\eta}\left(a^2\hat{q}_s \partial_{\eta}\tilde{\hat{\Psi}}_B\right) + c_S^2 \delta^{ij}k_ik_j \tilde{\hat{\Psi}}_B  = 0\,,
\end{eqnarray}
in terms of the conformal time $\eta$, where $\tilde{\hat{\Psi}}_B$ is the Fourier transformation of $\hat{\Psi}_B$. This is in the same form as Eq.(\ref{eq3.19aa}) for tensor perturbations. Therefore, we can follow the analysis given in the last subsection to carry out the analysis of the propagation of the scalar GWs in EsGB gravity.

Writing $\tilde{\hat\Psi}_B$  in the form 
\begin{equation}
\label{eq4.58}
    \tilde{\hat\Psi}_B = \hat{\cal{A}}_S e^{i\Theta_S/\epsilon}\,, 
\end{equation} 
 where $\Theta_S/\epsilon$ denotes the phase of the scalar GW, which  in general is different from that of the tensor modes introduced in the last subsection. From  Eq.(\ref{eq4.57}) we find 
\begin{eqnarray}
\label{eq5.59}
\epsilon^{-2}: \; \omega_S^2 - c_S^2 \delta^{ij}k_ik_j = 0\,,
\end{eqnarray}
and 
\begin{eqnarray}
\epsilon^{-1}: \; 2\omega_S\hat{\cal{A}}'_S + \omega'_S\hat{\cal{A}}_S + \omega_S\hat{\cal{A}}_S\partial_{\eta}\ln (a^2\hat{q}_S) = 0\,,
\end{eqnarray}
where $\omega_S\equiv -\partial_{\eta}\Theta_S$ and $k_i/\epsilon$ is the comoving wavevector. 

To proceed further, let us first show that $c_S \simeq 1$ in all epochs of the evolution of the Universe. To this goal, we first note that in the radiation-dominated epoch we have
\begin{eqnarray}
\alpha_S &=& \frac{1}{\tilde\Delta}\left(\frac{t_{Rc}}{t}\right)\,,\nonumber\\
\tilde\Delta &\equiv& \left(1 - \epsilon_R\right)  + \left(2 - \epsilon_R\right)\left(\frac{t_{Rc}}{t}\right)^{1/2}  \nonumber\\
&&  + \frac{11}{8}\left(\frac{t_{Rc}}{t}\right)\,, ~~~~~~~~~
\end{eqnarray}
where $t_{Rc}$ and $\epsilon_R$ are defined by Eqs.(\ref{eq2.4mm}) and (\ref{eq2.4oo}) respectively. From these definitions we find
\begin{eqnarray}
\epsilon_R \simeq 0.586, \quad \left(\frac{t_{Rc}}{t_{\text{reh}}}\right)^{1/2}
\lesssim 1.24 \times 10^{-21}\,.
\end{eqnarray}
Therefore, we find that
\begin{eqnarray}
\left|\alpha_S(t)\right| \lesssim {\cal{O}}\left(\frac{t_{Rc}}{t_{\text{reh}}}\right) 
\lesssim {\cal{O}}\left(10^{-42}\right)\,.
\end{eqnarray}

In the matter-dominated epoch, it can be shown that
\begin{eqnarray}
\alpha_S &=& \frac{1}{\tilde\Delta}\left(\frac{t_{Dc}}{t}\right)^{2/3}\,,\nonumber\\
\tilde\Delta &\equiv& \left(1 - \epsilon_D\right)  + \left(2 - \epsilon_D\right)\left(\frac{t_{Dc}}{t}\right)^{1/3} \nonumber\\
&& + \frac{3}{2}\left(\frac{t_{Dc}}{t}\right)^{2/3}\,,
~~~~~~~
\end{eqnarray}
where $t_{Dc}$ and $\epsilon_D$ are defined by Eqs.(\ref{eq2.4ss}) and (\ref{eq2.4vv}) respectively. From these definitions we find
\begin{eqnarray}
\epsilon_D \simeq 0.655\,, \quad \left(\frac{t_{Rc}}{t_{\text{eq}}}\right)^{1/3}\,.
\lesssim  10^{-44}.
\end{eqnarray}
Hence, we obtain
\begin{eqnarray}
\left|\alpha_S(t)\right| \lesssim {\cal{O}}\left(\frac{t_{Dc}}{t_{\text{eq}}}\right)^{2/3} 
\lesssim {\cal{O}}\left(10^{-88}\right)\,.
\end{eqnarray}
On the other hand, in the $\Lambda$-dominated epoch, we have 
\begin{eqnarray}
\left|\alpha_S(t)\right| \propto \epsilon_{H} \simeq 0\,. 
\end{eqnarray}

Therefore, in all epochs we have 
\begin{eqnarray}
\left|\alpha_S(t)\right| \ll 1, \quad c_S^2 = 1 - \alpha_S(t) \simeq 1\,.   
\end{eqnarray}
Then, Eq.(\ref{eq5.59}) yields
\begin{eqnarray}
 |\omega_S^2 - \delta^{ij}k_ik_j| \ll \delta^{ij}k_ik_j\,,
\end{eqnarray}
that is, the trajectories of the scalar modes can be well approximated by null geodesics, defined by
\begin{eqnarray}
\frac{dx^{\mu}(\lambda)}{d\lambda} = k^{\mu}\,.
\end{eqnarray}
Thus, following what we did in the last subsection, we find that 
\begin{eqnarray}
\label{eq5.68}
\partial_{\eta} \left(a\hat{\cal{A}}_S \sqrt{\hat{q}_s}\right)  = 0\,.
\end{eqnarray}
Eq.(\ref{eq5.68}) has the general solution
\begin{eqnarray}
\label{eq5.71}
\hat{\cal{A}}_S = \hat{\cal{A}}_S^{(e)}\left(\frac{{\cal{R}}_e}{\cal{R}}\right) 
\left(\frac{\hat{q}_{se}}{\hat{q}_s(t)}\right)^{1/2}\,,
\end{eqnarray}
where ${\cal{R}}_e \equiv {\cal{R}}(t_e) = r_e a(t_e)$ and $\hat{q}_{se} \equiv \hat{q}_s(t_e)$ with $t_e$ being the mission time of the scalar GWs. Then, from Eqs.(\ref{eq4.56}) and (\ref{eq4.58}) we find that
\begin{eqnarray}
\tilde{\Psi}_B = {\cal{A}}_S e^{i\Theta_S/\epsilon}\,,
\end{eqnarray}
where $\tilde{\Psi}_B$ is the Fourier transformation of $\Psi_B$, 
\begin{eqnarray}
 {\cal{A}}_S(t)  \equiv  \frac{\kappa H\left(1+ 6\hat\alpha a H\right)}{aq_T^2}\hat{\cal{A}}_S = {\cal{A}}_S^{(0)}\left(\frac{\cal{D}}{q_T^3}\right)^{1/2}\,,
\end{eqnarray}
and ${\cal{A}}_S^{(0)} \equiv \sqrt{\kappa q_{se}/2}\; a_e  \hat{\cal{A}}_S^{(e)}$.

In the radiation-dominated epoch, $q_T$ and ${\cal{D}}$ are given by Eqs.(\ref{eq2.4mm}) and (\ref{eq2.4nnD}), from which we find that 
\begin{eqnarray}
 {\cal{A}}^{(R)}_S(t)   = \frac{{\cal{A}}_S^{(0)}}{\sqrt{2}\kappa}\left\{ 1 + 2 \left(\frac{t_{Rc}}{t}\right)^{1/2} + {\cal{O}}\left(\frac{t_{Rc}}{t}\right)\right\}\,, ~~~~
\end{eqnarray}
where $(t_{Rc}/t)^{1/2} \lesssim 1.24\times 10^{-21}$. Thus, if the scalar modes are produced during the radiation-dominated epoch, its amplitude remains almost constant. This is sharply in contrast to the tensor modes, which are always decaying as $1/{\cal{R}}$. 

In the matter-dominated epoch, $q_T$ and ${\cal{D}}$ are given by Eqs.(\ref{eq2.4ss}) and (\ref{eq2.4tt}), from which we find that 
\begin{eqnarray}
 {\cal{A}}^{(D)}_S(t)   &=& \frac{{\cal{A}}_S^{(0)}}{\sqrt{2}\kappa}\left\{ 1 + 2 \left(\frac{t_{Dc}}{t}\right)^{1/3} \right.\nonumber\\ 
 && ~~~~~~~~~ \left. + {\cal{O}}\left(\left(\frac{t_{Dc}}{t}\right)^{2/3}\right)\right\}\,, 
\end{eqnarray}
where $(t_{Dc}/t)^{1/3} \lesssim  10^{-44}$. Thus, similarly to that in the radiation-dominated epoch, now the amplitude of the scalar modes  also remains almost constant.

On the other hand, in the $\Lambda$-dominated epoch, we find 
\begin{eqnarray}
&& \dot{H}, \; \dot\phi \simeq 0, \quad  {\cal{D}} = 48\hat\alpha^2H_0^4 a^2\,,\nonumber\\
&& q_T = 1 + 4\hat\alpha H_0 a\,,
\end{eqnarray}
where $\hat\alpha H_0 \lesssim 6.8\times 10^{-47}$. Then, we obtain
\begin{eqnarray}
\label{eq5.72}
 {\cal{A}}^{(\Lambda)}_S(t)   &=& {\cal{A}}^{(\Lambda, 0)}_S 
 \Big(1  + {\cal{O}}\left(\hat\alpha H_0\right)\Big)\left(\hat\alpha H_0 a\right)\nonumber\\
 &\simeq &  {\cal{A}}^{(\Lambda, e)}_S\left(\frac{{\cal{R}}(t)}{{\cal{R}}_e}\right)\,, 
 \end{eqnarray}
 where ${\cal{A}}^{(\Lambda, 0)}_S \equiv \sqrt{48} H_0 {\cal{A}}_S^{(0)}$, ${\cal{R}}_e \equiv {\cal{R}}(t_e)$, and ${\cal{A}}^{(\Lambda, e)}_S$ denotes the initial amplitude of the GW emitted at $t = t_e$, which vanishes when $\alpha = 0$, as it is expected. This is because when $\alpha = 0$ EsGB gravity reduces to GR, in which scalar GWs do not exist.

The result of Eq.(\ref{eq5.72}) is unexpected, as it tells us that the amplitude gets increasing with the physical distance ${\cal{R}}$, instead of decreasing as $1/{\cal{R}}$, as that of the spin-2 GWs. A more careful analysis shows that this is due to the assumption  $c_T = 1$. In the following section, we shall show explicitly that this is no longer the case when $c_T \not= 1$, although we still keep the observational constraint (\ref{CD}) to hold.

\section{Spin-0 Gravitational Waves Produced in the DE-dominated epoch with $c_T \not= 1$}
\renewcommand{\theequation}{4.\arabic{equation}} \setcounter{equation}{0}

To consider GWs produced in the DE-dominated epoch with $c_T \not= 1$, let us first note that the stability conditions (\ref{eq2.4gg}) can be written as
\begin{eqnarray}
\left\{\left|\alpha H \dot{f}\right|, \left|\alpha \ddot{f}\right|\,, 
\left|\alpha H^2 f_{,\phi}\right|\right\} \ll 1\,.
\end{eqnarray}
Setting 
\begin{eqnarray}
c_T^2 = 1 + \alpha_T, \quad \alpha_T \equiv \frac{\alpha\left(\ddot{f} - H\dot{f}\right)}{q_T}\,,  
\end{eqnarray}
we find that the condition (\ref{CD}) now reads
\begin{eqnarray}
\label{eq5.3}
- 6\times 10^{-15} \lesssim \alpha_T \lesssim 1.4 \times 10^{-15}\,.  
\end{eqnarray}

It can be shown that the gauge-invariant quantity $\Psi_B$ satisfies the same equation as that given by Eq.(\ref{eq4.52}) but now with
\begin{eqnarray}
\label{eq5.4}
I &\equiv& \frac{\Delta}{2\kappa q_T}, \quad J \equiv \frac{H}{q_T}\left(1 + 6 \hat\alpha aH\right), \quad K\equiv \frac{q_T}{\kappa}\,,\nonumber\\
\Delta &\equiv& q_T\dot\phi^2 + 48 \alpha^2 H^4\dot{f}^2\left[1 - \frac{1}{3}\left(4\epsilon_H + \alpha_T\right)\right]\,,\nonumber\\
c_S^2 &\equiv& 1 - \alpha_S, \quad q_T = 1 +  4\alpha H\dot{f}\,, \nonumber\\
\alpha_S &\equiv&  \frac{48\alpha^2H^4\dot{f}^2}{q_T \dot{\phi}^2 + 48\alpha^2H^4\dot{f}^2}\left(\epsilon_H + \alpha_T\right)\,.  
\end{eqnarray}
Introducing $\hat{\Psi}_B$ as that given by Eq.(\ref{eq4.56}), we find that Eq.(\ref{eq4.52}) can be cast in the form
\begin{eqnarray}
\frac{1}{a^2 Q_S}\partial_{\eta}\left(a^2 Q_S \partial_{\eta}\tilde{\hat{\Psi}}_B\right) + c_S^2\delta^{ij}k_ik_j \tilde{\hat{\Psi}}_B  = 0\,,
\end{eqnarray}
where 
\begin{eqnarray}
Q_S \equiv \frac{J^2}{\kappa a^4 I} = \frac{2H^2\left(1 + 6 \alpha H \dot{f}\right)^2}{q_T \Delta a^4}\,.  
\end{eqnarray}

In the DE-dominated epoch, we have $\dot{\phi} \simeq 0 \simeq \dot{H}$. Then, from 
Eq.(\ref{eq5.4}) we find
\begin{eqnarray}
\alpha_S &\simeq \alpha_T \quad \Rightarrow \quad   \left|\alpha_S\right| \lesssim 10^{-15}\,. 
\end{eqnarray}
Thus, we have
\begin{eqnarray}
c_S = \sqrt{1 - \alpha_S} =  1  + {\cal{O}}\left(10^{-15}\right)\,.
\end{eqnarray}
Therefore, the trajectories of the spin-0 gravitons can still be well approximated by null geodesics. As a result, the amplitude of $\hat{\Psi}_B$ will take the same form as that of Eq.(\ref{eq5.71}) by replacing $\hat{q}_s$ by $Q_S$. Then, we find that
\begin{eqnarray}
\tilde{\Psi}_B = \frac{H\left(1+6\alpha H \dot{f}\right)}{a q_T^2} \tilde{\hat{\Psi}}_B = {\cal{A}}_S e^{i\Theta_S/\epsilon}\,,
\end{eqnarray}
where
\begin{eqnarray}
\label{eq5.10}
{\cal{A}}_S(t) = {\cal{A}}^{(e)}_s  \left(\frac{3 - 4\alpha_T(t)}{3 - 4\alpha_{Te}}\right)^{1/2} \left(\frac{q_{Te}}{q_T(t)}\right)^{3/2} \left(\frac{\dot{f}(t)}{\dot{f}(t_e)}\right)\,, \nonumber\\
\end{eqnarray}
$\alpha_{Te} \equiv \alpha_{T}(t_e)$, $q_{Te}  \equiv q_T(t_e)$, and ${\cal{A}}^{(e)}_s (\propto \alpha)$ is an integration constant denoting the initial amplitude of the scalar GWs. When $\alpha_T = 0$, we have $\alpha\dot{f} = \hat\alpha a$, and the above 
 expression reduces to that given by Eq.(\ref{eq5.72}). However, when 
\begin{eqnarray}
 |\alpha \ddot{f}| \ll 1\,, \quad  |\alpha H\ddot{f}| \ll 1\,,
\end{eqnarray}
the constraint (\ref{eq5.3}) can be easily satisfied without making $|\ddot{f}/(H\dot{f}) - 1|$ small. This enables us to evade the behavior $\alpha\dot{f}\propto a$ that would result in (\ref{eq5.72}). We thus conclude that the amplitude of the scalar GWs produced in the DE-dominated epoch is no longer proportional to ${\cal{R}}$, unlike Eq.(\ref{eq5.72}) that was obtained by assuming that $c_T = 1$ regorously holds.

\section{Conclusions}
\renewcommand{\theequation}{5.\arabic{equation}} \setcounter{equation}{0}

 In this paper, in the framework of EsGB gravity we studied the propagation of GWs first produced by remote compact astrophysical sources and then propagating through cosmic distances before arriving at detectors in the Universe. A distinguished feature of the theory is that it contains  spin-0  gravitational modes, in addition to the spin-2 modes existing in GR. Thus, the investigation of the spin-0 modes is very interesting and important, as their observations in the current and forthcoming GW observations can place GR and EsGB gravity directly under tests. In addition,   the forthcoming space- and ground-based  detectors
 are able to detect GWs emitted from  binary systems as far as the redshift is about $z \simeq 100$ \cite{HE19}, 
which will result in a variety of profound scientific consequences. In particular, GWs propagating over such long cosmic distances will carry valuable information not only about their sources, but also about  the detail of the cosmological expansion and inhomogeneities of the universe, whereby a completely new window to explore the universe by using GWs is opened.

With the above motivations, we first  studied the stability conditions of the theory and its observational constraints, as in the literature some contradicted conclusions regarding to these questions often raise. In particular, in the case $c_T = 1$ we found  that the stability conditions hold in all epochs of the evolution of the flat universe,  provided that the condition (\ref{CD1}) holds, while the severest observational constraints  are given by  Eq.(\ref{eq2.4uu}). 

In Section III, we studied the propagation of the spin-0 and spin-2 gravitational modes in the flat homogeneous and isotropic universe. When $c_T = 1$, we were able to integrate their trajectories explicitly, so are their amplitudes, over the whole evolution of the universe, including the radiation-, matter- and dark energy-dominated epochs. From these explicit expressions, we found that the amplitude of the spin-2 GW is indistinguishable from that of GR, while the spin-0 GWs remain almost constant during radiation- and matter-dominated epochs, and in the dark energy-dominated epoch it is proportional to the physical distance between the source and observer. 

In Section IV, we studied the propagation of the spin-0 graviton carefully without imposing the condition $c_T = 1$ in the dark energy-dominated epoch, but still assumed that the stability conditions and the observational constraints are satisfied. Then, we showed that the above growing behavior of the amplitude of the spin-$0$ GWs no longer holds, and instead the amplitude of the spin-0 GWs is given by Eq.(\ref{eq5.10}), which is proportional to $\dot{f}(t) \left(= f_{,\phi} \dot{\phi}\right) \ll 1$.
 
 It would be very interesting to study the effects of the inhomogeneities of the universe on the propagation of both spin-0 and spin-2 modes, and then comparing them with those obtained in GR \cite{Fier21,LLSY10}. We hope to return this topic soon.

 \begin{acknowledgments}
 
K.L. is supported by the Brazilian agencies Fapesq-PB. S.M. is supported in part by Japan Society for the Promotion of Science (JSPS) KAKENHI Grant Number JP24K07017, as well as the World Premier International Research Center Initiative (WPI), MEXT, Japan. A.W. is partially supported by the US NSF grant: PHY-2308845. 

\end{acknowledgments}

\appendix
\section{Appendix A: The Homogeneous and isotropic Universe in EsGB gravity}
\renewcommand{\theequation}{A.\arabic{equation}} 
\setcounter{equation}{0}
\label{AppendixA}

In this Appendix, we shall provide a very brief introduction to the flat FLRW universe
in the framework of the EsGB gravity. The spacetime is described by the metric  
\begin{eqnarray}
ds^2 &\equiv& {\gamma}_{\mu\nu} dx^{\mu} dx^{\nu}\nonumber\\
&=& -dt^2+a^2(t)\left(dx^2+dy^2+dz^2\right)\,,
\end{eqnarray}
where $t$ is the cosmic time. It is often to use the conformal time $\eta$ defined by $d\eta = dt/a(t)$, so the metric takes the form,
\begin{eqnarray}
ds^2  
&=& a^2(\eta)\left(-d\eta^2+dx^2+dy^2+dz^2\right).
\end{eqnarray}
Then, the Ricci scalar and the Gauss-Bonnet combination are
\begin{eqnarray}
R^{(0)} = 6\left(2H^2 +\dot{H}\right)\,, \quad
{\cal{G}}^{(0)}= 24 H^2\left(H^2 + \dot{H}\right)\,, ~~~
\end{eqnarray} 
where $H \equiv  \dot{a}/a$ and an overdot represents derivative w.r.t. $t$. Then, the EsGB and Klein-Gordon equations with a perfect fluid yield 
\begin{eqnarray}
\label{eq3.14}
H^2 &=& \frac{1}{6}\left[2\kappa \rho + \frac{1}{2} \dot{\phi}^2 + V - 24\alpha H^3\dot{f}\right]\,,\\
\label{eq3.15}
\dot{H} &=& -\frac{1}{4}\Bigg[2\kappa \left(\rho + p\right) + \dot{\phi}^2 + 16\alpha H \dot{H} \dot{f} \nonumber\\
&& ~~~~~~~~ + 8\alpha \left(\ddot{f} - H \dot{f}\right) H^2\Bigg]\,,\\
\label{eq3.16}
\ddot{\phi} &+& 3 H \dot{\phi}  + V_{,\phi} =  \alpha {\cal{G}}^{(0)} f_{, \phi}\,, 
\end{eqnarray}
where $\rho$ and $p$ are respectively the energy density and pressure of the perfect fluid
\begin{eqnarray}
T_{\mu\nu}^{m} = \left(\rho + p\right)u_{\mu}u_{\nu} + p \gamma_{\mu\nu}\,,
\end{eqnarray} 
and $u_{\mu}dx^{\mu} = -dt$. Then, the conservation law $\nabla^{\nu}T^m_{\mu\nu} = 0$ leads to
\begin{eqnarray}
\label{eq3.17a}
\dot{\rho} + 3 H \left(\rho + p\right) = 0.  
\end{eqnarray}
It should be noted that  Eq.(\ref{eq3.17a}) is not independent and can be obtained from Eqs.(\ref{eq3.14}) - (\ref{eq3.16}). In fact, among the four equations, Eqs.(\ref{eq3.14}) - (\ref{eq3.16}) and (\ref{eq3.17a}), (\ref{eq3.16}) can be obtained from the other three as far as $\dot{\phi}\ne 0$. Similarly, (\ref{eq3.15})  can be obtained from the other three as far as $H\ne 0$. 

 In terms of the conformal time $\eta$, we find that the above field equations read
\begin{eqnarray}
{\cal{H}}^2 &=& \frac{1}{6}\left[2\kappa a^2\rho + \frac{1}{2} (\phi')^2 + a^2 V - \frac{24\alpha f' }{a^2} {\cal{H}}^3\right]\,,\\
\cal{H}'   &=& {\cal{H}}^2 -\frac{1}{4}\Bigg[2\kappa a^2 \left(\rho + p\right) + (\phi')^2 + \frac{16\alpha{\cal{H}}f' }{a^2}\left({\cal{H}}' - {\cal{H}}^2\right)    \nonumber\\
&& ~~~~~~~~~~~~ + \frac{8\alpha}{a^2} \left(f''  -2 {\cal{H}} f' \right){\cal{H}}^2\Bigg]\,,\\
\phi'' &+& 2 {\cal{H}} \phi'  + a^2 V_{,\phi} =  \alpha a^2 {\cal{G}}^{(0)} f_{, \phi}\,,\\
\rho' &+& 3{\cal{H}} \left(\rho + p\right) = 0\,.
\end{eqnarray}

\end{document}